\documentclass[12pt]{iopart}

\usepackage{bm}
\usepackage{graphicx}
\usepackage{color}

\begin{document}

\title{Structural, magnetic and dielectric properties in 3$d$-5$d$ based Sr$_2$FeIrO$_6$ thin films}

\author{K C Kharkwal$^1$, Roshan Kumar Patel$^1$, K Asokan$^2$ and A K Pramanik$^1$}
\address{$^1$School of Physical Sciences, Jawaharlal Nehru University, New Delhi - 110067, India}
\address{$^2$Materials Science Division, Inter University Accelerator Centre, New Delhi- 110 067, India.}

\eads{\mailto{akpramanik@mail.jnu.ac.in}}

\begin{abstract}
The structural, magnetic and dielectric properties have been investigated in 3$d$-5$d$ based double perovskite Sr$_2$FeIrO$_6$ thin films deposited by pulse laser deposition technique. To understand the effect of strain, epitaxial films are grown with varying thickness as well as on different substrates i.e., SrTiO$_3$ (100) and LaAlO$_3$ (100). The films with highest thickness are found to be more relaxed. Atomic force microscope images indicate all films are of good quality where grain sizes increase with increase in film thickness. X-ray absorption spectroscopy measurements indicate a Ir$^{5+}$ charge state in present films while providing a detailed picture of hybridization between Fe/Ir-$d$ and O-$p$ orbitals. The bulk antiferromagnetic transition is retained in films though the transition temperature shifts to higher temperature. Both dielectric constant ($\epsilon_r$) and loss ($\tan\delta$) show change around the magnetic ordering temperatures of bulk Sr$_2$FeIrO$_6$ indicating a close relation between dielectric and magnetic behaviors. A Maxwell-Wagner type relaxation is found to follow over whole frequency range down to low temperature in present film. On changing the substrate i.e., LaAlO$_3$ (100), the $\epsilon_r(T)$ and ($\tan\delta(T)$) show almost similar behavior but $\epsilon_r$ shows a higher value which is due to an increased strain coming from high mismatch of lattice parameters.   
\end{abstract}


\submitto{\JPCM}

\maketitle
\section{Introduction}
\label{sec:Introduction}
Iridates have drawn significant interest over last one decade because of their novel physical properties which arise mostly due to its complex oxidation states (Ir$^{4+}$, Ir$^{5+}$, etc.) and strong spin-orbit coupling (SOC) effect of Ir.\cite{Pesin, Kim, Wan} Within the octahedral environment of oxygen, the extended 5$d$ orbitals of Ir splits into $t_{2g}$ and $e_g$ states with a large gap between them. In presence of strong SOC, the low lying $t_{2g}$ state further splits into $J_{eff}$ = 3/2 and 1/2 states. In case of Ir$^{4+}$ (5$d^{5}$), the low-lying $J_{eff}$ = 3/2 quartet is fully filled while leaving one electron in $J_{eff}$ = 1/2 doublet. Even, a small electron correlation ($U$) effect in 5$d$ based Ir opens up a gap in $J_{eff}$ = 1/2 state, giving a realization of $J_{eff}$ = 1/2 Mott insulating state.\cite{Kim, Kim1} On the other hand, following this model a nonmagnetic ground state ($J_{eff}$ = 0) is expected in Ir$^{5+}$ with 5$d^{4}$ electronic configuration. This picture has given interesting physics in materials such as, Sr$_2$IrO$_4$, Ba$_2$IrO$_4$, Na$_2$IrO$_3$, Y$_2$Ir$_2$O$_7$, La$_2$(Zn,Mg)IrO$_6$, Sr$_3$Ir$_2$O$_7$, Y$_2$Ir$_2$O$_7$ etc.\cite{Kim1, Bhatti, JW, Bose, Cao, Kumar} While most of the recent studies have focused on Ir$^{4+}$ based physics, a little attention has been focused on Ir$^{5+}$ based one.\cite{Khal} Recent investigations have, however, revealed that the magnetic nature of Ir$^{5+}$ which is quite debated, showing an importance of underlying lattice distortion.\cite{Cao1,Dey,Page}

The 3$d$-5$d$ based double perovskites (DPs) have shown interesting properties not only because of their lattice, charge, spin and orbital degrees of freedom but also due to an exotic interplay between $U$ and SOC which show prominent effect in 3$d$ and 5$d$ elements, respectively.\cite{Zhang,Morrow,Serrate} The Sr$_2$FeIrO$_6$ is a prominent example of 3$d$-5$d$ based double perovskite where Fe$^{3+}$(3$d^5$) and Ir$^{3+}$(5$d^4$) are arranged alternatively in unit cell. The crystal structure of this material is reported with either monoclinic or triclinic type by different groups.\cite{Battle, Bufaical, Qasim, Khark} An antiferromagnetic (AFM) type magnetic transition with $T_N$ $\sim$ 120 K has been reported in previous studies in this material.\cite{Battle,Bufaical,Qasim} In our recent investigation we, however, have shown that this material has two AFM transitions; one a weak transition around 120 K and another prominent transition around 45 K.\cite{Khark} Similar two AFM transitions with different spin structures have earlier been shown in other 3$d$-5$d$ based DP system Sr$_2$FeOsO$_6$.\cite{Paul, Williams} The transition metal ions are assumed to be in Fe$^{3+}$ (3$d^5$) and Ir$^{5+}$(5$d^4$) ionic states where the former contributes to magnetism while the latter is non-magnetic.\cite{Battle, Bufaical, Qasim, Khark} Given that Fe$^{3+}$ (0.645 \AA) and Ir$^{5+}$ (0.57 \AA) have difference in ionic radii, a well behaved ordering of Fe and Ir ions is expected in present compound. Our previous study has, in deed, indicated an ordered state for Fe/Ir ions in Sr$_2$FeIrO$_6$.\cite{Khark}

Thin films of oxide materials are of recent investigation from the perspective of device applications where the strain engineering proves to be an effective tool to tune the properties of bulk materials. Further, the 2D confinement of electrons modifies the physics which has notable consequences on device applications.\cite{Locquet, Haeni, Choi, Du, Saloaro} Unusual magnetism, electronic properties, superconductivity and strain induced exchange bias  etc. are prominent examples of novel properties found in oxide thin films.\cite{Chakhalian, Hwang, Palai, Singh} A little work though has been done with iridate films in this direction.\cite{Nichols, Miao, Lapascu} Further, in recent times the AFM materials are being considered as future candidate for spintronic applications, the field which at present is mostly dominated by ferromagnetic materials. The combined features of AFM materials that robust against fluctuations in magnetic fields, devoid of stray fields, having very fast spin dynamics and high magnetoresistance make them ideal choice for spintronic devices. Henceforth, the investigations of AFM based thin films attach attention in this regard.\cite{Baltz} 

Double perovskites are, in general, being widely investigated for interesting dielectric behavior due to their high insulating nature.\cite{Yang,Jwang,Dutta} An exotic magneto-electrical coupling makes these materials suitable for technological applications.\cite{Gonz,Anshul} Interfaces in heterostructures further exhibit many exotic phenomena which mainly arisedue to various effects such as, strain from lattice mismatch, Rashba type spin-orbit coupling effect due to breaking of lattice inversion symmetry, induced potential difference due to electronic charge transfer and/or charge accumulation, oxygen non-stoichiometry, etc.\cite{Tokura,Zubko}  While 3$d$-3$d$ based DPs are mostly studied for their electrical investigations, the 3$d$-5$d$ based DPs needs to be explored to examine the combined effect of crystal structure, $U$ and SOC on dielectric properties.\cite{Yang,Jwang,Dutta,Gonz,Anshul} Here, it can be mentioned that oxidation state of transition metal ions play a crucial role in determining their physical properties.\cite{Hauser, Lee, Marco} Here, we have studied the crystal structure, surface morphology, magnetic and dielectric properties of Sr$_2$FeIrO$_6$ thin films with varying thickness where the strain is controlled using different substrates. Films are found to be epitaxial to the substrate where the grain size of films decreases with decrease in film thickness. The films show an AFM transition at low temperature where an interesting coupling between magnetic and dielectric properties are observed across the transition temperature.

\begin{figure*}
	\centering
		\includegraphics[width=15cm]{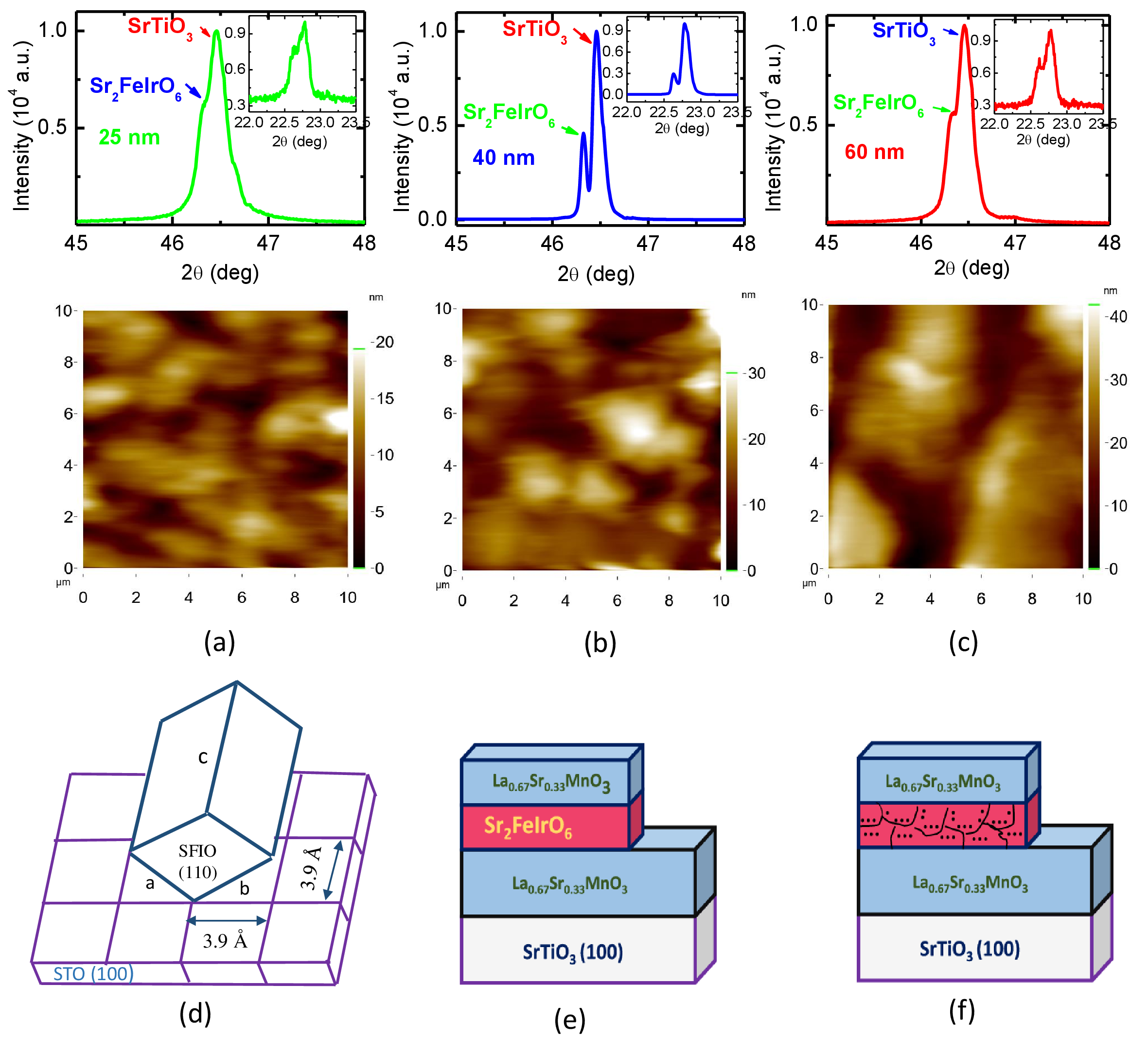}
	\caption{(Color online) (a), (b) and (c) show the x-ray diffraction data along with atomic force microscope images for 25, 40 and 60 nm thick Sr$_2$FeIrO$_6$ films, respectively which are grown on SrTiO$_3$ substrate. Insets of the figures show the XRD data at low angle regime with intensity of the order of $10^{3}$ a.u. The main panel shows the XRD Bragg peaks for (200) plane while that for (100) plane are shown in inset of respective figures. (d) shows schematic of film growth direction (110) along with substrate crystal direction (100). (e) shows the schematic diagram of film deposition for dielectric measurements where $\sim$ 25 nm thick Sr$_2$FeIrO$_6$ film is sandwiched between $\sim$ 15 nm La$_{0.67}$Sr$_{0.33}$MnO$_3$ films as deposited on SrTiO$_3$ (100) substrate. The La$_{0.67}$Sr$_{0.33}$MnO$_3$ films are used as top and bottom electrodes for dielectric measurements. (f) shows schematic diagram for accumulation of charge carriers at grain boundaries in Sr$_2$FeIrO$_6$ thin film.}
	\label{fig:Fig1}
\end{figure*}

\section{Experimental Methods}
\label{sec:Experimental Methods}
Single layer thin films of Sr$_2$FeIrO$_6$ have been grown on single crystal SrTiO$_3$ (100) substrate using pulsed laser deposition (PLD) technique equipped with KrF eximer laser ($\lambda = 248$ nm). The deposition of films has been done with substrate temperature of $700^{0}$C and oxygen pressure 0.1 mbar while laser energy density used as 1.3 J/cm$^2$. The thickness of the films have been kept around 60, 40, 25 nm, respectively. Thickness of films has been determined from laser shot calibration, details are given elsewhere.\cite{kk} For dielectric measurement first $\sim$ 15 nm thick La$_{0.67}$Sr$_{0.33}$MnO$_3$ (LSMO) film has been deposited on SrTiO$_3$ (100) substrate. The system is then cooled to room temperature, and a small mask has been used on a small portion of deposited LSMO film where subsequently $\sim$ 25 nm Sr$_2$FeIrO$_6$ and $\sim$ 15 nm LSMO films have been grown following above mentioned parameters (see Figure 1e). LSMO layers are used as electrodes which is known to be metallic below room temperature. To understand the effect of strain on dielectric properties, another $\sim$ 5 nm thick Sr$_2$FeIrO$_6$ film with similar configuration (see Figure 1e) has been grown on LaAlO$_3$ (100) substrate.

The present thin films are characterized using x-ray diffraction where the data have been collected with a PANalytical X'pert PRO instrument. This machine is equipped with Ge based monochromators (in both incident and diffracted beam optics) and Ni foil (in incident beam optics) which make x-ray radiation truly monochromatic with only Cu-$K_{\alpha1}$ source ($\lambda = 1.54$ \AA). Surface topography of the film has been checked using atomic force microscope (Nanomagnetics Instruments) operated in non-contact mode. Magnetization measurements have been done in VSM-SQUID magnetometer (Quantum Design). Dielectric measurements have been done in home made system attached with closed cycle refrigerator (CCR) where the data have been collected with LCR meter (NF Corporation, ZM2376). X-ray absorption (XAS) data for 40 nm Sr$_2$FeIrO$_6$ thin film grown on SrTiO$_3$ (100) substrate have been collected from $'$National synchrotron radiation research center$'$, Taiwan for O-$K$ edge and Ir-$L_{3}$ edge in total electron yield (TEY) and fluorescence modes, respectively by following standard procedure.

\section{Results and Discussions}
\subsection{Structural characterization and surface morphology}
In transition metal DPs ($A_2BB^/$O$_6$ where $A$ is alkaline- or rare-earth elements and B/B$^/$ are the transition metals), the transition metal octahedra (BO$_6$ and B$^/$O$_6$) are alternately arranged in two interpenetrating sublattices. Therefore, the crystal structure of related material immensely depends on the ionic sizes of $A$, $B$ and $B^/$ elements. While the tolerance factor (1.01) calculated using the ionic radii of constituent elements of bulk Sr$_2$FeIrO$_6$ indicates a cubic structure but the same calculated with real lattice parameters (from Rietveld analysis) comes out to be 0.95 which suggests a distorted lattice such as, monoclinic or triclinic structure. The crystal structure of bulk Sr$_2$FeIrO$_6$ is though debated showing it to be monoclinic-$P2_1/n$,\cite{Battle,Bufaical} monoclinic-$I2/m$,\cite{Qasim} or triclinic-$I\bar{1}$.\cite{Battle} However, we have previously shown a distorted triclinic-$I\bar{1}$ structure for bulk Sr$_2$FeIrO$_6$ with lattice constants $a$ = 5.5514, $b$ = 5.5784, $c$ = 7.8435 \AA, and angles $\alpha$ = 90.092, $\beta$ = 89.866 and $\gamma$ = 89.960 deg.\cite{Khark} The magnetic exchange interaction is dominated via Fe-O-O-Fe pathway in $ab$-plane and Fe-O-Ir-O-Fe pathway along $c$-axis within the unit cell as Ir$^{5+}$ (5$d^{4}$, $J_{eff}$ = 0) does not contribute to the magnetic interaction.\cite{Khark} Interesting properties are expected in thin films of Sr$_2$FeIrO$_6$ when the dimension is reduced in one of the lateral direction. The x-ray diffraction data along with atomic force microscope images for films with thickness around 25, 40, 60 nm are shown in Figures 1a, 1b, 1c, respectively. In main panel of figures, the XRD Bragg peaks for (200) plane while in inset the same for (100) plane have been shown. Thin films are found to be epitaxial to the substrate. Single crystalline peaks of films along with substrate in the XRD data indicate films take similar structure to the substrate. Surface morphology of grown films have been checked with atomic force microscope. The lower part of Figures 1a, 1b, 1c shows the atomic force microscope images of respective films. The images show an uniform growth with an average grain size around 2, 2.5 and 4 $\mu$m for 25, 40, 60 nm thick films, respectively. The root mean square (rms) roughness values for 25, 40, 60 nm thick films found to be around 3.54, 5.15, 5.48 nm, respectively, which demonstrate films are of good quality.

\begin{figure}
	\centering
		\includegraphics[width=7cm]{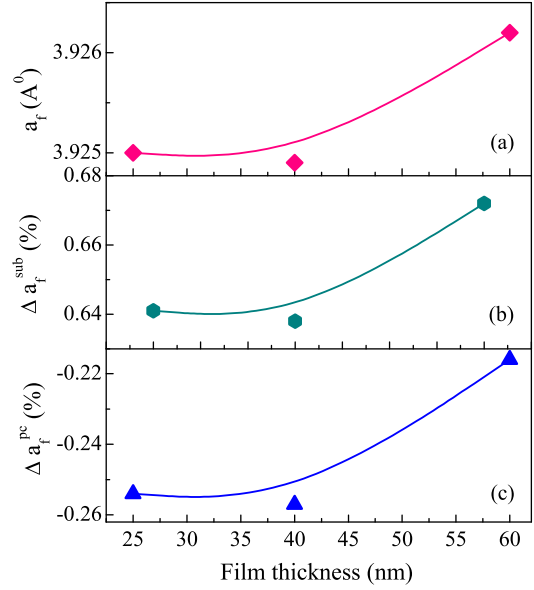}
	\caption{(Color online) The film thickness dependent (a) lattice parameter ($a_f$) as calculated form XRD data, (b) lattice mismatch between film and substrate (Eq. 1) and (c) lattice mismatch between film and bulk pseudo-cubic value (Eq. 2) are shown for Sr$_2$FeIrO$_6$ thin film grown on SrTiO$_3$ substrate.}
	\label{fig:Fig2}
\end{figure}
 
The lattice parameters of bulk Sr$_2$FeIrO$_6$ are $a$ = 5.5515, $b$ = 5.5785, $c$ = 7.8435 {\AA} whereas the lattice parameter of SrTiO$_{3}$(100) substrate $a_{sub}$ is 3.90 {\AA}. The out-of-plane growth of films takes (110) direction while $c$-axis of films lie in plane of the substrate. Figure 1d shows a schematic diagram for the growth direction of films. The pseudo-cubic parameter ($a_{pc}$) is calculated from bulk lattice parameters using the formula $\sqrt{(a^2 + b^2)}/2$, which gives $a_{pc}$ = 3.935 {\AA}. The lattice parameters of films ($a_f$) are calculated from the Bragg peaks in XRD data (Figures 1a, 1b and 1c) and have been shown as a function of film thickness in Figure 2a. A small decrease in $a_f$ has been found initially from 25 to 40 nm and then it increases for highest thick film (60 nm), however, the change is not substantial. This means with increasing film thickness the film lattice parameter $a_f$ moves towards its bulk $a_{pc}$ value (3.935 \AA). Further, to understand the lattice strain, the lattice mismatch ($\Delta a_f^{sub} \%$) between substrate and film has been calculated using following formula,

\begin{eqnarray}
\Delta a_f^{sub} \% = \frac{(a_{f} - a_{sub})}{a_{sub}} \times 100 \%
\end{eqnarray} 

Figure 2b shows the thickness variation of $\Delta a_f^{sub} \%$ showing the mismatch increases with the increase of film thickness. Considering that present films are epitaxial, therefore an increasing lattice mismatch $\Delta a_f^{sub} \%$ implies film relaxes with increasing thickness and lattice strain is released. However, the observed strain ($a_f$ $>$ $a_{sub}$) is compressive in nature and applies along the growth direction. 

Further, to understand the change in film lattice parameter $a_f$ with respect to its bulk pseudo-cubic value $a_{pc}$, the following $\Delta a_f^{pc} \%$ parameter has been calculated,

\begin{eqnarray}
\Delta a_f^{pc} \% = \frac{(a_{f} - a_{pc})}{a_{pc}} \times 100 \%
\end{eqnarray}

The calculated $\Delta a_f^{pc} \%$ as a function of film thickness has been shown in Figure 2c. The magnitude of $\Delta a_f^{pc} \%$ decreases with thickness which implies $a_f$ approaches $a_{pc}$ or lattice strain is released.

\subsection{X-ray absorption spectroscopy analysis} 
In order to understand the electronic state of Ir and the hybridization state of Fe/Ir-O in Sr$_2$FeIrO$_6$ thin films, x-ray absorption (XAS) measurements have been performed on Ir-$L_3$ and O-$K$ absorption edge, respectively. The $L_3$ absorption edge involves transition from 2$p_{3/2}$ to 5$d_{3/2}$ and 5$d_{5/2}$ states and usually occurs around 11220 eV.\cite{Kim,Marco,Clancy}. Figure 3 shows the XAS spectrum of Sr$_2$FeIrO$_6$ film for Ir-$L_3$ absorption edge where the position of edge agrees well with other Ir based compounds. The spectrum shows a broad peak around 11222 eV, indicating a large local density of 5$d$ state. In XAS spectra, in general, the position of absorption edge or peak is related to the ionic state or transition metal. For instance, the energy shifting of absorption edge is largely determined by the transition metal - oxygen bond length. With an increase in ionic state (shorter bond-length) the energy edge occurs at higher energy. However, a small difference in edge position has been observed even for same ionic state with different lattice structure. For Ir-$L_3$ absorption spectra, it has been previous seen that the peak position for Ir, Ir$^{3+}$, Ir$^{4+}$, Ir$^{5+}$ and Ir$^{6+}$ occurs around 11218.0, 11219.6, 11220.0, 11222.0 and 11222.5 ev, respectively.\cite{Kim,Marco,Clancy,Harish,Liu} In this sense, the Ir-$L_3$ peak position in Figure 3 ($\sim$ 11222 eV) for present film implies a Ir$^{5+}$ oxidation state in present film. It can be noted that the peak position of Ir-$L_3$ absorption edge in present Sr$_2$FeIrO$_6$ film matches with its bulk counterpart.\cite{Marco} Further, in inset of the Figure 3, we have shown the second derivative of Ir-$L_3$ absorption edge where a shoulder like feature (marked by * in figure) is observed at higher energy side of the peak where a similar shoulder has been observed in bulk Sr$_2$FeIrO$_6$ .\cite{Marco} While this implies a Ir$^{5+}$ oxidation state, further from separation between the main peak and shoulder, an estimate about the crystal field splitting can be made.\cite{Marco} The inset of Figure 3 shows the separation between shoulder and main peak in Ir-$L_3$ absorption edge is about 2 eV which is high and is expected for 5$d$ elements.  

\begin{figure}
	\centering
		\includegraphics[width=7cm]{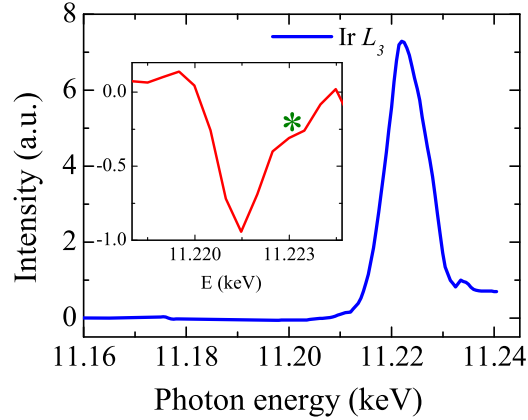}
	\caption{(Color online) The Ir-$L_3$ absorption edge x-ray absorption (XAS) data for Sr$_2$FeIrO$_6$ film with linearly subtracted pre-edge and normalized post-edge are shown. Inset shows second derivative of Ir-$L_3$ absorption edge where the asymmetric shoulder is shown with asterisk mark.}
	\label{fig:Fig3}
\end{figure}

\begin{figure}
	\centering
		\includegraphics[width=7cm]{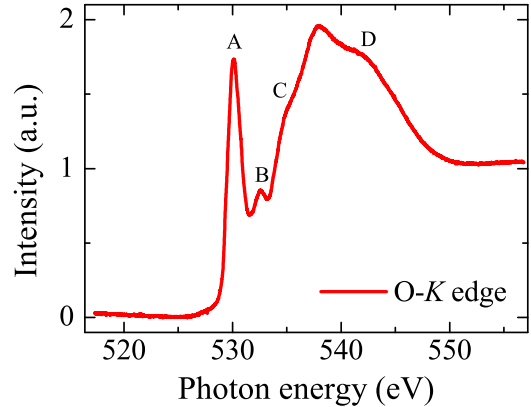}
	\caption{(Color online) The O-$K$ absorption edge x-ray absorption (XAS) data for Sr$_2$FeIrO$_6$ film with linearly subtracted pre-edge and normalized post edge are shown. A, B, C and D represents the peak positions.}
	\label{fig:Fig4}
\end{figure}   

In transition metal oxides, the metal-oxygen octahedra mainly govern the physical properties where the six oxygen (four basal and two apical) 2$p$ orbitals ($p_x$, $p_y$, $p_z$) hybridize with $d$ orbitals of transition metals (Fe/Ir). The $t_{2g}$ orbitals have $d_{xy}$, $d_{xz}$, $d_{yz}$ states whereas $e_{g}$ contains $d_{x^{2}-y^{2}}$ and $d_{z^2}$ states. For $t_{2g}$, the $d_{xy}$ hybridizes with basal $p_x$/$p_y$ while $d_{xz}$ and $d_{yz}$ hybridizes with both basal $p_z$ and apical $p_x$/$p_y$ orbitals of oxygen. The $d_{x^{2}-y^{2}}$ and $d_{z^2}$ of $e_g$, on the other hand, hybridize with basal $p_x$/$p_y$ and apical $p_z$ orbitals, respectively.\cite{Liu,Harish,Seong,Ilak} The O-$K$ edge XAS spectrum of Sr$_2$FeIrO$_6$ film is shown in Figure 4 for the energy range between 517 to 557 eV. A linear background has been subtracted from the XAS data by fitting in the lower energy range ($<$ 528 eV) and data are normalized to unity for high energy range ($>$ 550 eV). Two peaks with peak positions 530.0 and 532.46 eV (A and B, respectively in Figure 4) are observed at low energy side of the spectrum, while a broad hump is observed between 534 to 549 eV. The peak at 530 eV mainly arises due to the hybridization of Fe/Ir-$d_{xz}$/$d_{yz}$ orbital with apical O-$p_x$/$p_y$ orbitals while peak at 532.46 eV comes from the hybridization of Fe/Ir-$t_{2g}$ orbital with basal O-2$p$ orbitals.\cite{Liu,Harish,Seong,Ilak} The broad hump between 534 and 549 eV in the spectrum mainly comes from hybridization of Fe/Ir-$e_{g}$ orbitals with O-2$p$ orbitals. The XAS results in Figure 4 clearly shows the hybridization picture of $t_{2g}$/$e_g$ orbitals of transition metals with $p_x$, $p_y$ and $p_z$ of anion oxygen. It is, however, further noticed in Figure 4 that the broad hump in high energy side exhibits prominent shoulders (C and D in Figure 4). Here, it can be mentioned that above hybridization picture has been discussed considering a isolated model of $e_g$ and $t_{2g}$ orbitals. A recent theoretical model has, however, shown that in 5$d$ based oxides, the crystal field effect, SOC and $U$ has prominent role on mixing of $e_g$ and $t_{2g}$ orbitals, hence influences the absorption edge spectra.\cite{stam} Therefore, XAS spectra likely to differ slightly in case of bulk materials and films considering lattice strain plays vital role in films. Nonetheless, theoretical calculations may be helpful to understand the role of orbital mixing in giving rise to the shoulder in O-$K$ edge spectra.      

\begin{figure}
	\centering
		\includegraphics[width=7cm]{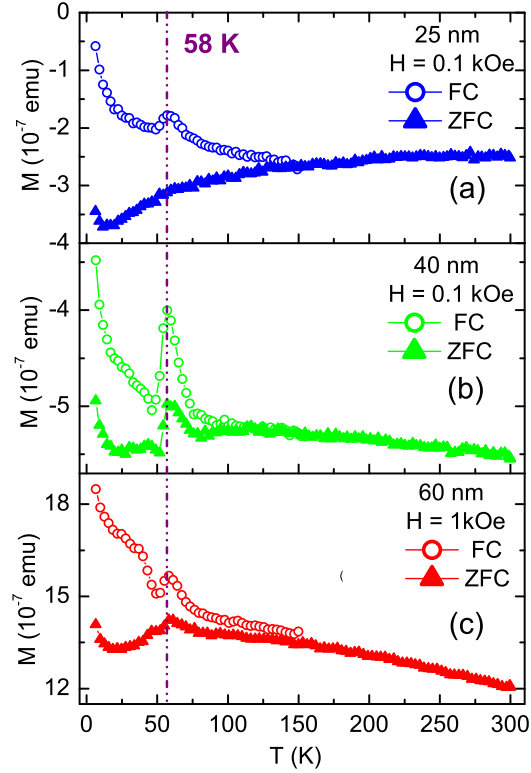}
	\caption{(Color online) Temperature dependent dc magnetization data collected following zero field cooled (ZFC) and field cooled (FC) protocol are shown for (a) 25 nm (b) 40 nm (c) 60 nm thich Sr$_2$FeIrO$_6$ film. The applied measurement fields are mentioned in respective plots.}
	\label{fig:Fig5}
\end{figure}

\subsection{Magnetic properties of Sr$_2$FeIrO$_6$ thin film}
Two AFM transitions, a sharp transition around 45 K and a weak transition around 120 K have been seen in bulk Sr$_2$FeIrO$_6$ magnetization data.\cite{Khark} The onset of AFM ordering appears to start around 120 K which becomes prominent at $\sim$ 45 K.\cite{Khark} The low temperature AFM ordering around 45 K is even preserved in magnetization data for (Sr$_2$FeIrO$_6$/La$_{0.67}$Sr$_{0.33}$MnO$_3$)$_3$ multilayer which has shown interesting exchange bias effect below 45 K.\cite{kk} The temperature dependent magnetization $M(T)$ data for single-layer thin films collected following zero field cooled (ZFC) and field cooled (FC) protocol have been shown in Figure 5. For collection of $M(T)$ data, the magnetic field has been used as 1000 Oe for 60 nm film (Figure 6c) and 100 Oe for 40, 25 nm thick film (Figures 6b and 6a). As evident in Figure 5, an AFM type magnetic transition is observed at $T_N$ $\sim$ 58 K for all the films. The magnetic transition temperature $T_N$ does not change much with the film thickness. Further, the moments are quite low which is mostly due an AFM type magnetic ordering as well as due to non-magnetic nature of Ir$^{5+}$ elements. Figures 5a and 5b show that for films of 25 and 40 nm thickness, the substrate diamagnetic contribution dominates giving a negative magnetization. However, a sense of magnetic transition can be obtained from the peaks in corresponding magnetic data. However, the data show that $T_N$ is shifted to high temperature compared to bulk Sr$_2$FeIrO$_6$. The shift in $T_N$ is understood due to strain effect arising from substrate. 

\begin{figure}
	\centering
		\includegraphics[width=7cm]{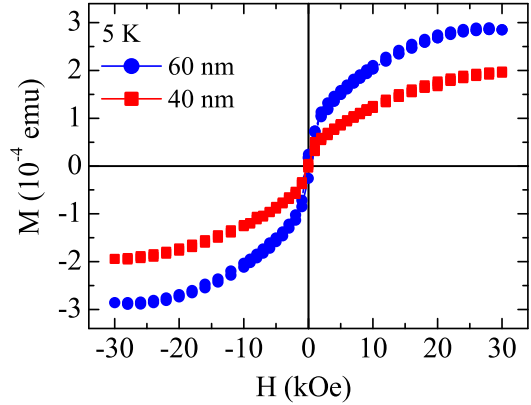}
	\caption{(Color online) The magnetic field dependent magnetization $M(H)$ data collected at 5 K are shown for 40, 60 nm thick Sr$_2$FeIrO$_6$ films.}
	\label{fig:Fig6}
\end{figure}

Field dependent magnetization $M(H)$ data collected at 5 K have been shown in Figure 6 for selected 60 and 40 nm thick Sr$_2$FeIrO$_6$ films. Substrate contributions are subtracted from measured $M(H)$ data by linear fit in high field regime. A small hysteresis has been found in the $M(H)$ data for both the films. The observed remnant magnetization ($M_r$) and coercive field ($H_c$) are around $2.5 \times 10^{-6}$, $3.9 \times 10^{-7}$ emu and 258, 90 Oe for 60, 40 nm thick films, respectively. $M(H)$ data show a sign of saturation at highest measured magnetic field i.e, around 30 kOe.

\subsection{Dielectric properties of Sr$_2$FeIrO$_6$ thin film}
Bulk Sr$_2$FeIrO$_6$ shows an highly insulating behavior down to low temperature where the resistivity increases to the value of around 10$^5$ $\Omega$-cm at 20 K from $\sim$ 20 $\Omega$-cm at room temperature.\cite{Khark} Sr$_2$FeIrO$_6$ thin films are also found to be highly insulating (resistance not measurable), therefore it would be interesting to study the dielectric properties of films. Dielectric measurements are usually done keeping the dielectric material between two parallel metallic plates. Figure 1e shows schematic configuration for measuring the dielectric properties in present films where an insulating layer of Sr$_2$FeIrO$_6$ is sandwiched between two conducting La$_{0.67}$Sr$_{0.33}$MnO$_3$ layers which has been used as top and bottom electrodes.

The dielectric constant ($\epsilon_{r}$) has been calculated from measured capacitance ($C$) using following formula, 

\begin{equation}
\epsilon_{r} = \frac{Cd}{\epsilon_{0}A}
\end{equation} 

where $d$ is distance between conducting plates, $A$ is the area of the plates and $\epsilon_{0}$ is the permittivity of free space.

Dielectric constant ($\epsilon_{r}$) and loss tangent ($\tan\delta$) have been shown in Figures 7a and 7b, respectively for Sr$_2$FeIrO$_6$ film grown on SrTiO$_3$(100) substrate at selected frequencies in temperature range 20-300 K. It is evident in Figure 7a that on heating, $\epsilon_{r}$ initially increases showing a peak at temperature $T_p$ and then decreases sharply above $T_p$. However, above $\sim$ 135 K, $\epsilon_{r}$ becomes almost constant. The $T_p$ is seen to be sensitive with applied frequency where the its value is found to be 35 K, 40 K and 45 K for 0.1, 1 and 10 kHz, respectively. Peak in $\epsilon_{r}$ is not very clear at higher frequencies where $\epsilon_{r}$ decreases continuously with increase in temperature. Loss tangent ($\tan\delta$) shows multiple relaxation in the present film (Figure 7b). With an increasing temperature, $\tan\delta$ is found to be almost constant till $T_p$. However above respective $T_p$, it increases sharply almost by ten times till about 135 K. A small decrease in $\tan\delta$ has been observed above 135 K with a peak around 280 K. Furthermore, $\tan\delta$ decreases monotonically with the increase in frequency. 

\begin{figure}
	\centering
		\includegraphics*[width=7cm]{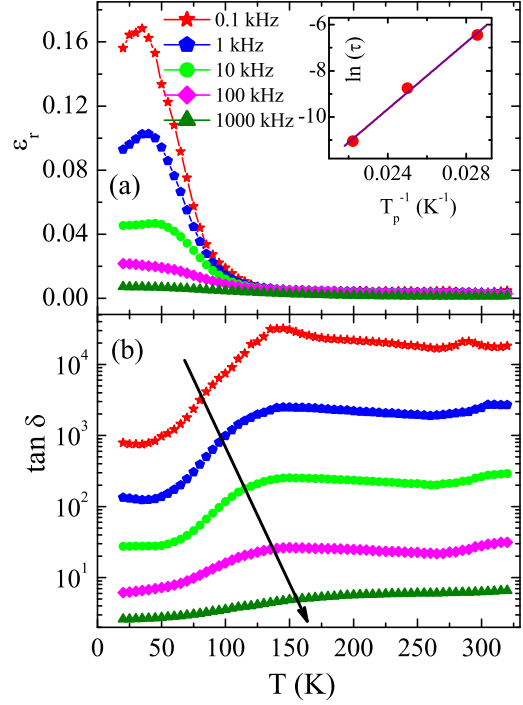}
	\caption{(Color online) (a) Dielectric constant ($\epsilon_{r}$) and (b) loss tangent ($\tan\delta$) are shown as a function of temperature for Sr$_2$FeIrO$_6$ thin film grown on SrTiO$_3$(100) substrate at various frequencies. Inset of (a) shows linear fit of relaxation time ($\tau$) and peak temperature ($T_p$) following Eq. 4.}
	\label{fig:Fig7}
\end{figure}

The overall behavior of $\epsilon_{r}$ and $\tan\delta$ at low temperature (Figures 7a and 7b) appears to be related with the magnetic behavior of Sr$_2$FeIrO$_6$. For instance, bulk Sr$_2$FeIrO$_6$ has double AFM magnetic transition  around temperatures 120 and 45 K.\cite{Khark} Even in Sr$_2$FeIrO$_6$ films, while the high temperature transition at 120 K is not clearly seen (which was itself weak in bulk material), the low temperature magnetic transition is retained and occurs around 58 K. Therefore, the change in $\epsilon_{r}$ and $\tan\delta$ in temperature window between 120-35 K is likely induced by spin ordering in film.   

The peak temperature ($T_p$) in $\epsilon_{r}$ shifts toward higher temperature with increase in frequency ($f$) indicating the relaxation mechanism is thermally activated. The relaxation mechanism at $T_p$ follows Arrhenius-type behavior which is given by following equation,

\begin{equation}
\tau = \tau_{0}\exp \left(\frac{E_a}{k{_B}T_p}\right)
\end{equation}     

where $\tau = 1/(2\pi f)$ is relaxation time at $T_p$ for frequency $f$, $\tau_{0}$ is characteristic time, $E_a$ and $k{_B}$ are the activation energy and Boltzmann constant, respectively. Inset of the Figure 7a shows a linear fit of relaxation time ($\tau$) following Eq. 4. An activation energy ($E_a$) from fitting has been found to be 0.62 eV which is comparable with the materials where the charge conduction happens through either hopping of electrons or oxygen vacancies.\cite{Shulman}  

To understand the nature of charge carriers in present film, ac conductivity ($\sigma_{ac}(T)$) has been calculated using temperature dependent $\epsilon_{r}$ and $\tan\delta$ data from Figure 7 with following relation,

\begin{eqnarray}
	 \sigma_{ac} = \epsilon_0 \epsilon_r \omega \tan\delta 
\end{eqnarray}

where $\epsilon_0 = 8.85 \times 10^{-12}$ F/m, $\omega = (2\pi f)$ and $f$ is frequency. The calculated $\sigma_{ac}(T)$ have been shown in Figure 8a at selected frequencies for Sr$_2$FeIrO$_6$ thin film grown on SrTiO$_3$(100) substrate. The $\sigma_{ac}(T)$ is of the order of $10^{-7}/\Omega$-m representing the value similar to highly insulating materials. On heating, $\sigma_{ac}(T)$ initially increases till around 120 K for 0.1, 1, 10, 100 kHz frequencies, however, a change in slope is observed across magnetic ordering temperature (45 K). Above 120 K, $\sigma_{ac}(T)$ data almost merge and then decrease till $\sim$ 275 K. With further rise in temperature, $\sigma_{ac}(T)$ again increases. The present film is found to be more conducting with increasing frequency i.e., 1000 kHz. This can be understood as accumulation of charge carriers at grain boundaries. Though films are epitaxial but a small inhomogeneity during growth, which is evident in atomic force microscope images (see Figures 1a, 1b and 1c), piles up charges at growth boundaries. Application of high frequency helps the charge carriers to cross the barrier and hop from boundaries which gives higher conduction in higher frequency. However, a dip in $\sigma_{ac}(T)$ around 275 K and an then increase of its value down to $\sim$ 120 K looks like the behavior of only La$_{0.67}$Sr$_{0.33}$MnO$_3$ film with similar thickness.\cite{Yuan1} The contribution in $\sigma_{ac}(T)$ below 120 K is mainly dominated by Sr$_2$FeIrO$_6$ where $\sigma_{ac}(T)$ increases with temperature presenting an insulating system. Nonetheless, an increase of $\sigma_{ac}(T)$ with frequency at low temperature imply an accumulation of loose charges at boundaries and and its assisted conduction with applied frequency.

\begin{figure}
	\centering
		\includegraphics*[width=7cm]{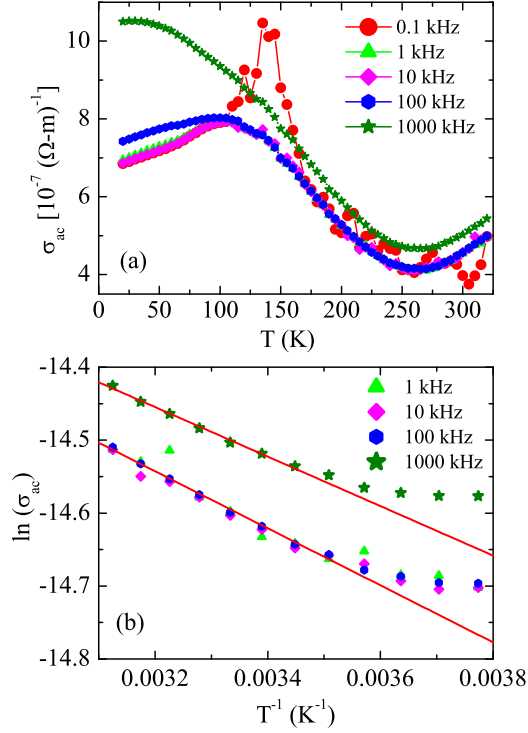}
	\caption{(Color online) (a) Temperature dependent ac conductivity $\sigma_{ac}(T)$ are shown for Sr$_2$FeIrO$_6$ thin film  grown on SrTiO$_3$(100) substrate at selected frequencies. (b) shows semi-log plotting between ac conductivity and inverse temperature for the same film. Solid lines are due to straight line fit to the data following Eq 6.}
	\label{fig:Fig8}
\end{figure} 

At high temperature intrinsic contribution due to charge carriers dominates, therefore $\sigma_{ac}(T)$ data has been analyzed using following equation,

\begin{eqnarray}
	\sigma_{ac} = \sigma_{0} \exp(-\frac{E_{a}}{k_{B} T})
\end{eqnarray}

Figure 8b shows fitting of $\sigma_{ac}(T)$ data following Eq 6. An activation energy ($E_a$) has been obtained from fitting of the data. $E_a$ is 0.29 and 0.34 eV for 1000 kHz and other frequencies, respectively. These value of $E_a$ implies thermally activated hopping mechanism of charge carriers in the film.

\begin{figure}
	\centering
		\includegraphics[width=7cm]{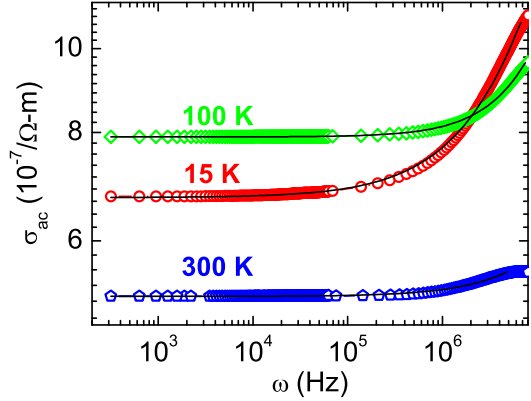}
	\caption{(Color online) Frequency dependent ac conductivity ($\sigma_{ac}$) has been shown on semi-log plot at selected temperatures for Sr$_2$FeIrO$_6$ thin film grown on SrTiO$_3$(100) substrate. Solid lines are due to fitting of the data following Eq. 7.}
	\label{fig:Fig9}
\end{figure}

It is evident from previous discussions that $\sigma_{ac}(T)$ is quite sensitive to both applied frequency as well as magnetic ordering temperature of Sr$_2$FeIrO$_6$. Following these, a frequency dependent $\sigma_{ac}(\omega)$ has been calculated (in range 0.1 - 3000 kHz) from frequency dependent $\epsilon_{r}$ and $\tan\delta$ using Eq 5. Figure 9 shows $\sigma_{ac}(\omega)$ for Sr$_2$FeIrO$_6$ thin film grown on SrTiO$_3$(100) substrate at few selected temperatures which represent magnetic (15 and 100 K) and nonmagnetic (300 K) state of present DP material. The $\sigma_{ac} (\omega)$ is found to be almost constant till about 100 kHz, and then increases at higher frequencies. Further, in frequency range below $\sim$ 10$^6$ Hz, $\sigma_{ac}$ increases with temperature within magnetic state (15 and 100 K) while showing a lower value in nonmagnetic state at 300 K. At higher frequencies ($>$ 10$^6$ Hz) there is a crossover in $\sigma_{ac}$ where $\sigma_{ac}$ collected at low temperature (15 K) shows a faster increase which is in line with the results shown in Figure 8a. This implies that the charge carriers which are trapped at low temperatures show higher hopping rate at higher frequencies. The $\sigma_{ac} (\omega)$ data have been fitted with following Jonscher's universal power law:\cite{Jons} 

\begin{eqnarray}
	\sigma_{ac} = \sigma_{dc} + A \omega^{n}
\end{eqnarray}

where $\sigma_{dc}$ is the dc conductivity, $A$ is the temperature dependent constant which describes the strength of polarizability of material and $n$ is the power exponent which describes the nature of interaction between mobile ions and surrounding lattice. Usually, $n$ takes value 1 for non-interacting Debye type systems while a smaller value of $n$ implies an interaction between mobile ions with the lattice. The solid lines in Figure 9 are due to fitting of $\sigma_{ac}(\omega)$ data with Eq. 7. The $\sigma_{dc}$, $A$ and $n$ obtained from fitting of $\sigma_{ac}(\omega)$ are shown in Table 1. From Table 1, it is evident that the change in $\sigma_{dc}$ with temperatures follows similar trend of $\sigma_{ac}(T)$ (Figure 9) i.e., conductivity increases with temperature within the magnetic state. Similarly, the value of $n$ indicates that system becomes non-interactive at higher temperature within magnetic state. The value of $A$ is found to be quite small indicating a weak polarization strength.

\begin{table}
\caption{\label{tab:table 1} DC conductivity ($\sigma_{dc}$),  power exponent ($n$) and pre-factor $A$ obtained from fitting of $\sigma_{ac}(\omega)$ data with Eq. 7 (Figure 9) are shown at different temperatures.}
\begin{tabular}{cccc}
\hline
$T(K)$ & $\sigma_{dc}$ ($10^{-7}/\Omega-m$) & $n$  & A ($10^{-13}$)\\
\hline
15 &  6.73 & 0.75 & 31.2\\
100 &  7.91 & 1.01 & 0.19\\
300 &  5.18 & 0.82 & 1.11\\ 
\hline
\end{tabular}
\end{table}
 
In order to get further insight about the charge conduction mechanism, a Cole-Cole plot is constructed in terms of electrical modulus for Sr$_2$FeIrO$_6$ film grown on SrTiO$_3$(100) substrate. The real ($M'$) and imaginary part ($M''$) of electrical modulus have been calculated using following relations,

\begin{eqnarray}
M' = \frac{\epsilon_{r}}{{\epsilon_{r}}^{2}+ {\epsilon''}^{2}}\\
M'' = \frac{\epsilon''}{{\epsilon_{r}}^{2}+ {\epsilon''}^{2}}
\end{eqnarray}

where $\epsilon'' = \epsilon_{r}\tan\delta$. The benefit of electrical modulus analysis is that it suppresses the electrode polarization and provides an information about the intrinsic conduction mechanism of mobile ions within the system.\cite{Yang,Jwang,Dutta} Electrical modulus analysis even identifies a minute capacitance in the system. Figure 10 shows the Nyquist plot in terms of electrical modulus ($M''$ vs $M'$) for selected temperatures at 15, 100, 300 K. The $M''$ vs $M'$ data approaches towards zero at low frequencies for all temperatures, indicating a suppression of electrode capacitance in the present thin film. Presence of asymmetric semi-circle further indicates multiple relaxation in the system. Two semi-circles are evident at each temperature (Figure 10). The semi-circle at low frequency side (small $M'$) represents grain boundary contribution while that at higher frequency side is due to contribution of grains within the film. It is known that the capacitance is inversely proportional to the radius of the semi-circle. As evident in Figure 10, both grain as well as grain boundary radius decreases with decrease in temperature. Figure 10 further shows that grain boundaries have smaller radius compared to grains at all temperatures. These imply the grain boundaries have more contribution to the system capacitance compared to grains, and with decrease in temperature the contribution of brain boundaries increases. This increase of capacitance with decrease of temperature (smaller radius) agrees with increasing dielectric constant value ($\epsilon_r$ $\propto$ $C$, Eq. 3) at low temperature, as shown in Figure 7a. 

\begin{figure}
	\centering
		\includegraphics[width=8cm]{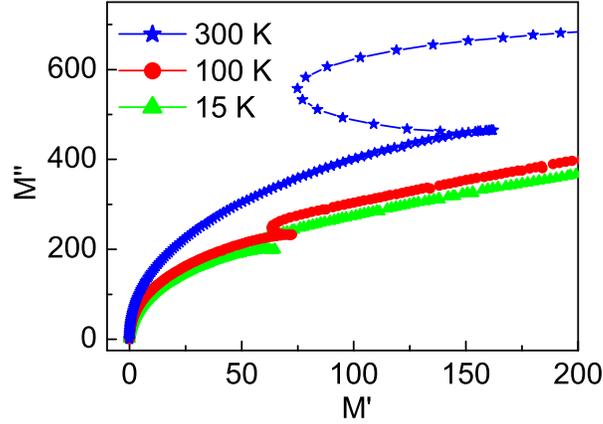}
	\caption{(Color online) Nyquist plot in terms of real ($M'$) and imaginary ($M''$) part of electrical modulus are shown at 15, 100, 300 K for Sr$_2$FeIrO$_6$ thin film grown on SrTiO$_3$ substrate.}
	\label{fig:Fig10}
\end{figure}

Two independent mechanisms, in general, contribute to the dielectric relaxation in a system. First one is dipolar contribution which mainly arises due to an asymmetric hopping of charge carriers (i.e., Mn$^{+3}$-Mn$^{+4}$, Fe$^{+3}$-Ir$^{+5}$ in the present film). The another one arises due to an accumulation of charge carriers in regions within material that have different conductivity such as, grain boundaries, interfaces, etc which hop to other regions. While the dipolar contribution provides Debye-type relaxation,\cite{Kao} the relaxation originating from hopping of accumulated charge carriers are known as Maxwell-Wagner (MW) type one.\cite{Hipp} At a particular temperature, with the help of applied frequency the accumulated charge carriers cross the barrier leading to relaxation in the inter-facial (space-charge) polarization.\cite{Hipp} The MW-type relaxation mechanism can further be identified from the inverse frequency ($f^{-1}$) dependence of imaginary part of dielectric constant ($\epsilon''$).\cite{Hipp} In order to elucidate the relaxation mechanism in present film, log-log plot of $\epsilon''$ and angular frequency ($\omega$) has been analyzed which is shown in Figure 11 for selected 15, 100, 300 K. Slopes obtained from the linear fittings of data are found to be around -1 for all temperature where this inverse frequency dependence of $\epsilon''$ implies MW-type relaxations are active in the present film over the whole frequency range down to low temperature.

\begin{figure}
	\centering
		\includegraphics[width=7cm]{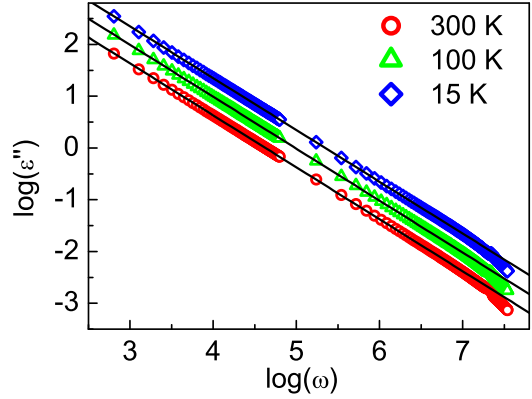}
	\caption{(Color online) A $\log-\log$ plot between imaginary part of dielectric constant ($\epsilon''$) and angular frequency ($\omega$) are shown at 15, 100, 300 K for Sr$_2$FeIrO$_6$ thin film grown on SrTiO$_3$(100) substrate. Solid lines are linear fit to the data showing a MW-type relaxation model in present film.}
	\label{fig:Fig11}
\end{figure} 

The value of $\epsilon_{r}$ is less than one in present film (Figure 7a) is though unusual. Usually, in an applied electric field, the effective induced polarization aligns in opposite direction of applied field which gives dielectric constant higher than one. The $\epsilon_{r}$ $<$ 1 in present film is likely due to the fact that fraction of polarization orient in same direction of electric field. With an increase of frequency, more number of polarization orient in direction of field which causes further decrease in $\epsilon_{r}$ value (Figure 7a). The induced polarization in present film can microscopically be thought of arising from deformation of ionic pairs such as, Mn$^{3+}$-O-Mn$^{4+}$, Fe$^{3+}$-O-Fe$^{3+}$, Fe$^{3+}$-O-Ir$^{5+}$ within the respective layers and Fe$^{3+}$-O-Mn$^{3+/4+}$, Ir$^{5+}$-O-Mn$^{3+/4+}$ at interface. 

The $\tan\delta$ basically signifies loss contribution in dielectric data which arises from local induced polarization as well as long range conduction. The long range contribution can be seen in $\sigma_{ac}(T)$ (Figure 10a) which is more conducting for 1000 kHz compared to other frequency data which are found to merge with separation around magnetic ordering temperature. Therefore long-range conduction gives almost similar change in loss data for all frequencies.  Contacts, which are made up of silver on top and bottom layers of La$_{0.67}$Sr$_{0.33}$MnO$_3$ will significantly increase the loss. However, induced negative polarization in present film is the cause of large loss.\cite{Yang, Jwang, Dutta} Close to room temperature only La$_{0.67}$Sr$_{0.33}$MnO$_3$ orders magnetically. Therefore, peak like feature observed at $\sim$ 280 K in the loss data arises from this ordering where relaxation mechanism dominates via Mn$^{3+}$-O-Mn$^{4+}$.

\subsection{Effect of strain in dielectric properties of Sr$_2$FeIrO$_6$ thin film}
In order to understand the effect of strain on dielectric properties, another Sr$_2$FeIrO$_6$ film with similar configuration (Figure 1e) has been grown on LaAlO$_3$(100) substrate. The pseudo-cubic lattice constant of LaAlO$_3$(100) is 3.79 \AA. The lattice mismatch between film of Sr$_2$FeIrO$_6$ and LaAlO$_3$(100) substrate is about -3.83$\%$ which is higher than the films grown on SrTiO$_3$(100) substrate. This large mismatch is expected to induce more strain in film and modify the dielectric properties accordingly.

\begin{figure}
	\centering
		\includegraphics*[width=7cm]{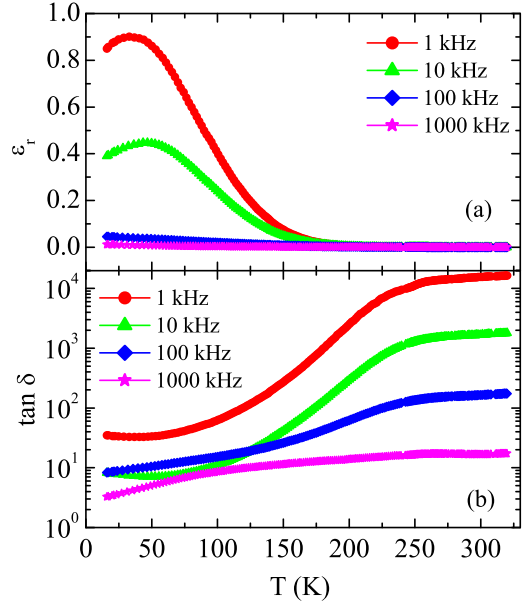}
	\caption{(Color online) (a) Dielectric constant ($\epsilon_{r}$) and (b) loss tangent ($\tan\delta$) are shown as a function of temperature for Sr$_2$FeIrO$_6$ thin film grown on LaAlO$_3$(100) substrate at selected frequencies.}
	\label{fig:Fig12}
\end{figure}

Temperature dependent dielectric constant ($\epsilon_{r}$) and loss tangent ($\tan\delta$) at selected frequencies have been shown in Figures 12a and 12b, respectively or Sr$_2$FeIrO$_6$ film grown on LaAlO$_3$(100) substrate. It is evident from Figure 12 that dielectric behavior is somewhat different than the film grown on SrTiO$_3$(100) substrate (Figure 7). Although the low frequency data of $\epsilon_{r}$ exhibit peak around magnetic ordering temperature ($\sim$ 45 K) for both films deposited on both SrTiO$_3$ and LaAlO$_3$ substrates but the film grown on LaAlO$_3$(100) shows higher $\epsilon_{r}$ at respective frequencies which is likely due to an enhanced strain effect with LaAlO$_3$(100) substrate. With increasing frequency, $\epsilon_{r}$ decreases monotonically but above $\sim$ 200 K, $\epsilon_{r}$ becomes insensitive to frequency showing nearly a constant value. The value of $\epsilon_{r}$ are however less than one which again indicates fraction of induced polarization is in line with applied electric field in the film. 

On heating, $\tan\delta$ data is almost constant till about 45 K (Figure 12b). Above 45 K, an increase in $\tan\delta$ is observed till about 250 K and then it becomes almost constant. Change in dielectric data around 250 K and 45 K suggest thermally activated mechanism of relaxation in the film. Furthermore, increasing the frequency $\tan\delta$ decreases monotonically.

\begin{figure}
	\centering
		\includegraphics*[width=7cm]{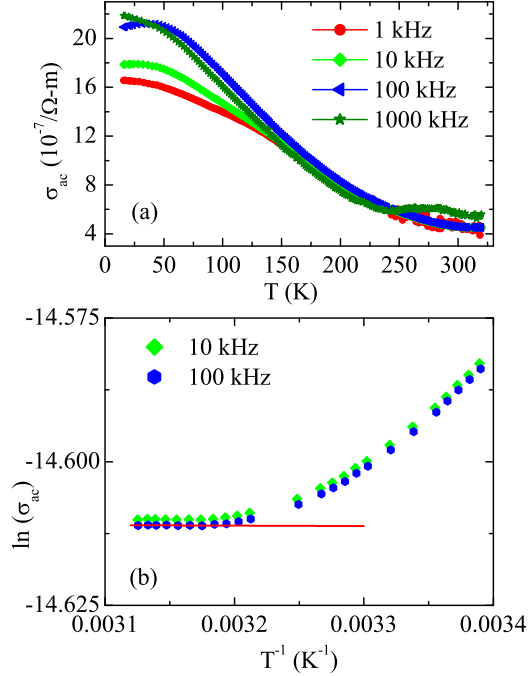}
	\caption{(Color online) (a) Ac conductivity ($\sigma_{ac}$) data calculated using Eq. 5 are shown as a function of temperature for Sr$_2$FeIrO$_6$ thin film grown on LaAlO$_3$(100) substrate at selected frequencies. (b) shows semi-log plotting of ac conductivity ($\sigma_{ac}$) and inverse temperature for Sr$_2$FeIrO$_6$ thin film grown on LaAlO$_3$(100) substrate at selected frequencies. Solid line is representative straight line fitting to the data following Eq. 6.}
	\label{fig:Fig13}
\end{figure}

The ac conductivity ($\sigma_{ac}$) has further been calculated for Sr$_2$FeIrO$_6$ thin film grown on LaAlO$_3$(100) using Eq. 5. The Figure 13a shows temperature dependence of $\sigma_{ac}$ at few selected frequencies. On heating, $\sigma_{ac}(T)$ mainly decreases though an anomaly is observed across its magnetic transition temperature (45 K). The $\sigma_{ac}(T)$ data for all frequencies merge above 150 K and then decreases with further rise in temperature. $\sigma_{ac}(T)$ is of the order of 10$^{-7}/\Omega-m$ which is similar to those of film grown on SrTiO$_3$(100) substrate (Figure 8a). Variation of $\sigma_{ac}(T)$ with temperature is almost similar to film grown on SrTiO$_3$(100) substrate (Figure 10a).

Similar to Figure 8b, $\ln \sigma_{ac}$ vs T$^{-1}$ has been plotted in Figure 13b at 10 and 100 kHz frequencies for Sr$_2$FeIrO$_6$ thin film grown on LaAlO$_3$(100). As evident in figure, at high temperature regime a linear behavior is obtained. An activation energy ($E_a$) has been obtained from the slope of fitting (Eq. 6) which turns out to be about 0.07 meV. Although in high temperature regime the charge conduction is mostly influenced by top and bottom layer of La$_{0.67}$Sr$_{0.33}$MnO$_3$ electrodes, but for the film with LaAlO$_3$(100) substrate shows comparatively very low activation energy.

\section{Conclusion}
In summary, epitaxial films of Sr$_2$FeIrO$_6$ have been grown on SrTiO$_3$ (100) and LaAlO$_3$ (100) substrates. Atomic force microscope image shows good quality of films with an increase of grain size with increase in film thickness. X-ray absorption spectroscopy measurements on Ir-$L_3$ edge reveals 5+ oxidation state for Ir while O-$K$ edge shows a hybridization between Fe/Ir-$d$ and O-$p$ orbitals. The magnetization data show that the bulk antiferromagnetic transition still occurs in present films but the transition temperature shifts to higher temperature which is understood due to strain effect. Both dielectric constant and loss show changes across the magnetic transition temperatures which imply a relation between magnetic and dielectric behavior. The activation energy calculated from the peak temperature of dielectric constant has been found to be 0.62 eV. Dispersion to high frequency has been found in the frequency dependent ac conductivity data which follows Jonscher's universal power-law behavior. Change in exponent $n$ (Table I) with temperature implies both non-interacting and interacting model of interaction between mobile ions and lattice. Further, a Maxwell-Wagner type dielectric relaxation has been found active in present film down to low temperature. The analysis of electrical modulus indicates grain boundaries contribute to the estimated capacitance at larger proportion compared to the grains. With the change in substrate i.e., LaAlO$_3$ (100) the dielectric properties changes significantly.  

\section{Acknowledgment}
\label{sec:Acknowledgment}
We thank Indian Institute of Science, Bangalore and Prof. P. S. Anil Kumar for the magnetic measurements. We are thankful to AIRF, JNU for the XRD measurements. We acknowledge SERB, DST for funding `Excimer Pulse Laser' and UPE-II, UGC for funding `Deposition Chamber' and `LCR Meter'. We also acknowledge DST-FIST for funding `Atomic Force Microscope' and DST-PURSE for the financial support. KCK acknowledges University Grant Commission, India for the financial support and KA thanks NSRRC, Taiwan for beamtime and support.

\end{document}